\documentstyle[preprint,aps,eqsecnum]{revtex}
\begin{document}
\preprint{PITT-94-05}
\draft
\title{\bf POLARONS AS \\
 NUCLEATION DROPLETS IN NON-DEGENERATE POLYMERS}
\author{\bf D. Boyanovsky$^{a)}$, C. A. A. de Carvalho$^{b)}$ and
 E. S. Fraga$^{c)}$}
\address{$^{a)}$ Department of Physics and Astronomy,
 University of Pittsburgh,\\
Pittsburgh, P.A. 15260, U.S.A. \\
$^{b)}$ Instituto de F\'\i sica, Universidade Federal
do Rio de Janeiro \\ C.P. 68528, Rio de Janeiro, RJ, 21945-970, Brasil \\
$^{c)}$ Departamento de F\'\i sica, Pontif\'\i cia Universidade Cat\'olica do
Rio de Janeiro\\ C.P. 38071, Rio de Janeiro, RJ, 22453-900, Brasil}
\date{}
\maketitle
\begin{abstract}
We present a study of the nucleation mechanism that allows the
decay of the metastable phase (trans-cisoid) to the stable phase
 (cis-transoid)
 in quasi one-dimensional non-degenerate polymers
within the continuum electron-phonon model. The electron-phonon
 configurations
that lead to the decay, i.e.  the critical droplets (or transition
state), are identified as
 polarons of the metastable phase.
 We obtain an estimate for the decay rate via thermal
activation within a range of parameters consistent with experimental
values for the gap of the cis-configuration. It is pointed out that,
upon doping, the activation barriers of the excited states are quite
smaller and the decay rate is greatly enhanced. Typical activation
energies for electron or hole polarons are $\approx 0.1$ eV
 and the typical size for a critical
droplet (polaron) is about $20 \AA$. Decay via quantum nucleation is
also studied and it is found that the crossover temperature
between quantum nucleation and thermal activation is of order
$T_c \leq 40 ^oK$. Metastable configurations of non-degenerate
polymers may provide examples for mesoscopic quantum tunneling.
\end{abstract}
\pacs{pacs No:71.10.+x;71.20.Hk;64.60.My}

\section{Introduction and Motivation}
Quasi-one dimensional  conductors provide a fascinating wealth of
physical phenomena that stimulated considerable theoretical and
experimental study of these systems. The proposal\cite{ssh} that solitons
 play a central role in the electronic and transport
 properties of the
degenerate polymer {\it trans}-$(CH)_x$ polyacetylene was later
confirmed with measurements of the optical absorption in the doped
system\cite{suz,sch}.
Non-topological excitations corresponding to local lattice deformations
that bind electron states, i.e. polarons, had also
 been recognized in these degenerate
polymers\cite{sus,bk}.

Although the presence of topological solitons is associated with the
ground state degeneracy of {\it trans}-polyacetylene, polarons are
not a consequence of this degeneracy and are predicted to occur upon
doping also in non-degenerate isomers\cite{bk}.

Non-degenerate isomers like {\it cis}-$(CH)_x$ were studied\cite{bk} by
using a continuum electron-phonon model\cite{tlm}.
These non-degenerate isomers feature a globally stable (lower energy)
configuration (cis-transoid) and a locally stable but globally metastable
(higher energy) configuration (trans-cisoid).

The properties of the polaron excitations of the {\it stable} (cis) phase
were thoroughly studied within the continuum electron-phonon model and
show the property of ``confinement'', that is their energy grows linearly
with the spatial size of the lattice distortion\cite{bk,fbc,martino,ylu}.
 Because of the
lack of degeneracy, there are no stable topological soliton excitations
in this case.

The motivation for our study is to understand the mechanism by which
the metastable configuration (trans-cisoid) decays into the stable phase
(cis-transoid). In particular, we identify the electron-phonon
 configurations that mediate the nucleation process as polarons of the
 {\it metastable phase}. These polaron configurations are
 the equivalent of
nucleation droplets, they correspond to saddle-points of the energy
functional and the critical droplets (critical radius of the polarons)
are identified with the transition state.
The ``radius'' of the polaron (droplet) is identified as the reaction
coordinate and eventually quantized as a collective coordinate.

 Once these configurations are identified, we
estimate the activation barriers and the decay rates via
thermal activation and investigate the possibility that the metastable
phase decays via quantum tunneling (quantum nucleation)
 at very low temperatures. Upon quantization of the
relevant collective coordinates
that describe the droplet-polaron, we provide a WKB estimate for
the rate of quantum nucleation.  Understanding
the relevant electron-phonon configurations that mediate the metastable
decay  allows us to
 provide an estimate for the crossover temperature between quantum
nucleation and thermal activation.

Despite a very large body of theoretical and experimental
 work on quasi-one dimensional conductors
 both
degenerate and non-degenerate, there has not been a consistent study
of the fundamental aspects of the decay of the metastable phase.

 A particularly interesting and tantalizing possibility that we explore
in this article is that the decay of the metastable phase may provide
an example of mesoscopic quantum tunneling.

The article is organized as follows: section II reviews the continuum
model for non-degenerate polymers. In section III we study the constant
dimerization configuration and establish the range of dimerization
values available, scanning through values of the intrinsic dimerization
consistent with experimental values of the gap in the stable (cis)
phase. The energy and dimerization values obtained in this section are
used to obtain the polaron excitations of the metastable phase.
Section IV is devoted to the study of the electron-phonon configurations
that mediate the nucleation. We construct the polaron solutions of the
{\it metastable phase}; these are nucleating droplets, whose energy as
a function of the ``radius'' of the droplet, which is identified as
the ``reaction coordinate'', has a barrier and is thus identified with a
saddle point in function space.
 The dependence of the energy and activation
barriers upon doping (occupation of electron bound states) is analyzed in
detail.
In section V we quantize the droplet configuration via collective
coordinates, obtain an estimate for the decay rate via thermal
activation and discuss the possibility of decay induced by doping
(electrons or holes). In section VI we study the decay of the metastable
phase via quantum tunneling using a WKB approximation for the collective
coordinates of the polaron. We estimate the crossover temperature between
thermal activation and quantum nucleation. Section VII summarizes our
results and presents our conclusions and speculations.

\section{The model Hamiltonian}

The {\it cis-transoid} isomer has a slightly lower energy than
the {\it trans-cisoid} isomer which is metastable. Because of the
slightly different energy of the two configurations, there is an
`intrinsic' dimerization that explicitly breaks the degeneracy
 between the
ground state configurations. As a result of this explicit symmetry
breaking, kink excitations are not available.

We consider the generalization of the SSH\cite{ssh} model Hamiltonian
 that
includes the non-degenerate case. The discrete version of the
 electron-phonon Hamiltonian has been introduced and discussed by
Wang, Su and Martino\cite{martino}, Kivelson\cite{kivelson1} and
 Lu\cite{ylu} and
is given by (here we assume the simpler case of a unique elastic constant
for the lattice distortion)

\begin{eqnarray}
 H & = &  -\sum_{n;s} \left\{[t_1-\alpha_1(U_{2n-1}-U_{2n})]
(C^{\dagger}_{2n,s}C_{2n-1,s}+h.c.)+ \right. \nonumber \\
   &   & \left. [t_2-\alpha_2(U_{2n}-U_{2n+1})](C^{\dagger}_{2n+1,s}C_{2n,s}
+h.c.)] \right\} +\nonumber\\
   &   & \sum_{n} \frac{K}{2} (U_{2n+1}-U_{2n})^2 +
 \frac{M}{2}\dot{U}^2_n
 \label{discreteham}
\end{eqnarray}

It proves convenient to introduce the parameters
\begin{eqnarray}
t_o & = & (t_1+t_2)/2 \; \; \; ; \; \; \; \alpha_o= (\alpha_1+
\alpha_2)/2 \label{alfa0t0}  \\
\Delta_e
    & = & (t_1-t_2) \; \; \; ; \; \; \; \Delta \alpha = (\alpha_1-
\alpha_2) \label{deltatalfa}
\end{eqnarray}

for the case $\Delta t =0 ; \Delta \alpha =0$ one recovers the original
SSH Hamiltonian for {\it trans-transoid}, which is the degenerate isomer.
Following the steps that led to the continuum version of the SSH model,
as described by Takayama, Lin-Liu and Maki\cite{tlm}, one is led to the
continuum version suggested by Brazovskii and Kirova\cite{bk}. As usual,
since only electron states near the Fermi surface are important and the
relevant phonon processes leading to the Peierls instability
 involve momenta of the order $2k_F$, the Fermi spectrum is linearized
near the two Fermi points, leading to a spinor description (left and
right branches). The lattice displacement is  written as

\begin{equation}
U_{n}= (-1)^n\left(\frac{\Delta(x)}{4\alpha_o}\right) \label{delta}
\end{equation}
where the `gap parameter' $\Delta(x)$ is a slowly varying function (on
the scale of a lattice spacing).
The continuum Hamiltonian, as derived by Brazovskii and Kirova\cite{bk},
becomes
\begin{eqnarray}
& H & =  \int dx \left\{\frac{M}{32\alpha_o^2 a}\dot{\Delta}^2(x)+
\frac{K}{8\alpha_o^2 a}\Delta^2(x)+ \right.
\nonumber \\
&   & \left. \sum_s\left(\Psi_s^{\dagger}(x)
\left[(-i)(v_F+\frac{a \Delta \alpha}{2\alpha_o}\Delta(x))\sigma_3
\frac{\partial}{\partial x}+(\Delta(x)+\Delta_e)\sigma_1\right]\Psi_s(x)
\right) \right\} \label{contham} \\  \nonumber \\
v_F & = & 2t_oa \label{fermivelocity}
\end{eqnarray}
where $a$ is the lattice spacing $a \approx 1.22 \AA$,
$\sigma_i$ are the Pauli matrices and $\Psi(x)$ is a  spinor. The label
$s=1,2$ corresponds to the two spin projections and plays a passive role.

For slowly varying $\Delta(x)$
(on the scale of a lattice spacing)  and small $\Delta \alpha$, the
term proportional to $\Delta \alpha$ is a small renormalization of
the Fermi velocity and of the same order as terms that have been
neglected in the derivative expansion leading to the continuum limit.
Thus, following the arguments presented by Lu\cite{ylu}, we will neglect
this term. The Hamiltonian obtained is the same as that considered
by Fesser, Bishop and Campbell\cite{fbc} as a model for
 $\rm{cis}-(CH)_x$.

Introducing the dimensionless electron-phonon coupling constant
 $\lambda$ and the bare
phonon frequency $\omega_Q$  as
\begin{equation}
\frac{K}{8\alpha_o^2 a}=\frac{1}{\pi v_F \lambda} \; \; ; \; \;
\omega_Q^2 =\frac{4K}{M}
\end{equation}
the model Hamiltonian for the non-degenerate isomer becomes
\begin{eqnarray}
H  & = & \int dx \left\{\frac{\dot{\Delta}^2(x)}{\omega_Q^2 \pi v_F \lambda}+
\frac{\Delta^2(x)}{\pi v_F \lambda}+ \right. \nonumber \\
   &   & \left. \sum_s
\Psi_s^{\dagger}(x)
\left[(-i)v_F\sigma_3
\frac{\partial}{\partial x}+(\Delta(x)+\Delta_e)\sigma_1\right]\Psi_s(x)
\right\} \label{finham}
\end{eqnarray}

We will concentrate, for the moment, on static configurations.
Following Wang, Su and Martino\cite{martino} and  Lu\cite{ylu} we will
{\it assume} the same value of the parameters as for the $trans-(CH)_x$
degenerate case as proposed by Su, Schrieffer and Heeger\cite{ssh}
\begin{eqnarray}
a  & = & 1.22 \AA \; \; \; ; \; \; \; t_o = 2.5 eV \label{paremeters1} \\
\alpha_o
   & = & 4.1 \frac{eV}{\AA} \; \; \; ; \; \; \; K = 21 \frac{eV}{\AA^2}
\label{parameters2}
\end{eqnarray}
leading to the following values of electron-phonon coupling $\lambda$,
band-width $W$, Fermi velocity $v_F$ and phonon frequency
\begin{equation}
\lambda  =   0.4077 \; \; ; \; \;
W       = 4t_o = 10 \rm{\mbox{ eV }} \; \; ; \; \;
v_F      =   2t_oa = 6.10 \rm{\mbox{ eV }} \AA \; \; ; \; \;
 \omega_Q \approx 0.14 \rm{\mbox{ eV }}
\label{parameters3}
\end{equation}
in units in which $\hbar=1$.  Since values of the intrinsic
dimerization are not experimentally available (see discussions in
\cite{martino,ylu}) we will search for a range of values for
$\Delta_e$ such as to reproduce the value for the energy gap for
$\rm{cis}-(CH)_x$ (which is the lower energy stable ground state
 configuration); this value is approximately 2.05 eV\cite{martino,ylu}.

\section{Constant Dimerization}

The constant dimerization case corresponds to assuming
$\Delta(x)= \Delta_o$ (a space independent value). The fermions have
a constant ``mass'' given by
\begin{equation}
M_f = \Delta_o + \Delta_e \label{fermionmass}
\end{equation}
In the Born-Oppenheimer approximation, the electronic energy is given
by the energy of the completely filled valence band when the fermions
have the above ``mass''
\begin{equation}
E_F= -2 \frac{L}{2\pi}\int^{k_c}_{-k_c}dk \sqrt{k^2v_F^2+
M_f^2}
\end{equation}
with $k_c$, the momentum cuttoff, related to the band-width $W$ as
\[W \approx 2k_c v_F \]
Incorporating the elastic energy, we obtain the
energy per site
\begin{equation}
\frac{E}{N} = \frac{a}{\pi v_F}\left[\frac{\Delta_o^2}{\lambda}
-2\int_0^{W/2}dz \sqrt{z^2+M_f^2} \right] \label{consdimE}
\end{equation}
The extrema of this energy function are determined by the
``gap equation''
\begin{equation}
\frac{\Delta_o^2}{\lambda} =
\int_0^{W/2}dz \frac{M_f}{\sqrt{z^2+M_f^2}}
\label{gapequation}
\end{equation}
For the values of the parameters given by Su, Schrieffer and Heeger
for the degenerate case (trans), we show in Figure (1) the
energy per site (in eV) as a function of $\Delta_o$ (in eV) for the
representative
value $\Delta_e= 0.02 \rm{\mbox{ eV }}$. As the value of $\Delta_e$ is
 increased, the stable minimum becomes deeper and the metastable minimum
 becomes shallower and closer to the origin.
When $\Delta_e \geq 0.07$(eV)
 the metastable minimum disappears altogether.
Numerically, the optimum range of $\Delta_e$ that
predicts a gap for the continuum theory with approximately the same
error as the continuum theory prediction for the degenerate case, with
the SSH values for the parameters (about 10 $\%$)\cite{note}, is $0.02 \leq
\Delta_e \leq 0.07$.
In this range we find:
\begin{equation}
\begin{array}{|r|c|c|c|l|}\hline
\Delta_e (eV) &  \Delta_{o-} (eV) & \frac{E_-}{N}(eV) & \Delta_{o+}(eV) &
\frac{E_+}{N}(eV) \\ \hline
0.02          & -0.835            & -1.610            & 0.896           &
-1.621 \\ \hline \label{tableone}
0.04          & -0.800            & -1.604            & 0.923           &
-1.626 \\ \hline
0.06          & -0.760            & -1.596            & 0.947           &
-1.632 \\ \hline
\end{array}
\end{equation}

Where $\Delta_{o\pm}$ correspond to the values of $\Delta_o$ at
 the metastable (-) and stable (+) minima with energy per site
 $E_{\pm}/N$
 respectively.

Thus, the value of the  gap  predicted from the global minimum
 ($2\Delta_{o+}$) for the stable configuration is fairly close to the
 observed gap 2.05 eV for
the lower energy $\rm{cis}-(CH)_x$ configuration (again the discrepancy is of
the
same order as the discrepancy between the gap predicted by the continuum
model and the observed value in the degenerate case (trans), with the
 value
of the parameters chosen by SSH\cite{note}). For values of
 $0.02 (eV) \leq \Delta_e \leq 0.07 (eV)$ curves very similar to figure
(1) are obtained for the energy per site.

\section{\bf Polaron Solutions}

The non-degenerate case does not permit soliton (kink) type excitations
(except if a very special theoretical possibility is fulfilled \cite{cac}),
but allows the possibility of polaron (bag) excitations. Polarons are
topologically trivial electron-phonon field configurations that correspond
to distortions of the lattice in which there are trapped electrons
(bound states). Because of charge conjugation symmetry, bound states must
appear symmetrically with respect to the middle of the gap.

For polaron solutions, the phonon field profile reaches asymptotically
the values of the minima of the effective energy for the constant
dimerization case, that is the minima in figure 1.  The polaron
solutions corresponding to the global (stable) minimum, had already
been studied in references(\cite{bk,fbc,ylu,cac}), these solutions
reach $\Delta_{o+}$  asymptotically when $|x| \rightarrow \infty$.
These solutions are
parametrized by {\it two} collective coordinates: the center of mass
of the polaron (position), associated with translational invariance, and
the radius. For the polaron solutions in the stable (lower energy) phase,
the polaron energy is a linearly increasing function of the radius, for
large radius. These polarons are ``confined''. The reason is simple:
for large radius, the polaron is exploring a region in function space
that is very close to the higher (metastable) minimum, thus increasing
the ``volume'' energy as a linear function of the radius.

In this article, we are interested in the polaron configurations in
the metastable phase. We argue that these are the relevant
electron-phonon configurations responsible for the {\it decay} of
the metastable state.

The argument is as follows: let us {\it assume} that a polaron solution
exists in the metastable phase. This configuration will
reach the values of the metastable minimum for the constant
 dimerization case $(\Delta_{o-})$ asymptotically as $|x| \rightarrow
\infty$. Its profile again will be that of a ``bag'', inside
which the field configuration will sample the lower energy minimum in
a region in space given by the radius of the polaron. However, the stable
minimum has lower energy and the system gains volume energy by increasing the
radius. There is a price in elastic energy determined
 by the gradients of the field configuration, but in one dimension
this ``surface energy'' is independent of the radius for large radius
 and bound by a value close to
twice the mass of the kink-antikink pair that is the asymptotic state
of the polaron for infinite radius. For small radius, the elastic term
wins out and the energy is an increasing function of the radius; for
large radius, the elastic energy will saturate and the volume energy
will dominate; in this case the energy is a decreasing function of
the radius. Thus, {\it if a polaron solution} exists with asymptotic values for
the dimerization in the metastable phase, there must be
a critical radius. This configuration is  equivalent to the  critical
``droplet''
configuration that Langer introduced to explain the statistical decay
of a metastable state in first order phase transitions\cite{langer} and
the ``radius'' of the droplet is identified with the reaction coordinate.

We will argue below that these polaron configurations will be saddle points of
the energy
functionals as they will be characterized by an unstable mode, a
zero mode and fluctuations with positive frequencies.

To actually construct the polaron configurations, we could invoke the
large body of work on the exact integrability of the continuum
theory \cite{bk,ylu,cb,dhn,cc,cac}, but we prefer to go through some of the
details of the calculation, as there are some subtle but important
features that must be addressed.

Borrowing from the known results for the degenerate case\cite{cb}
 and for the stable minimum\cite{bk,fbc,ylu}(cis) we propose
 the static polaron solution
 as
\begin{equation}
\Delta_p(x;x_{cm};x_o)=\Delta_o -K_o
v_f\left\{\tanh\left[K_o(x-x_{cm}+x_o)\right]-
\tanh\left[K_o(x-x_{cm}-x_o)\right]\right\} \label{polaron}
\end{equation}

The fermionic potential becomes reflectionless for the integrability
condition\cite{cb,dhn,cc,cac}

\begin{eqnarray}
\tanh\left[2K_ox_o\right] & = &  \frac{K_ov_f}{M_f}
\label{reflectionless} \\
             M_f          & = & \Delta_o+\Delta_e \label{deltatilde}
\end{eqnarray}

In the expressions above, $\Delta_o$ can be either one of
 the minima of the
energy functional for the constant dimerization case, i.e., $\Delta_{o\pm}$.

Thus, the polaron profile is parametrized in terms of $x_o$
and $x_{cm}$, the ``center of mass''; later we will treat these as
 ``collective coordinates''. The ``radius'' of the polaron is $2|x_o|$.
 The center of mass coordinate
$x_{cm}$ reflects the underlying translational invariance, the energy is
 independent of this coordinate.

Before proceeding further to the computation of the polaron energy, it
proves illuminating to understand some features of the polaron solution
and the integrability condition: i) the polaron solution (and
 consequently the
electronic spectrum) is an even function of $K_o$. We choose
$K_o >0$; ii) for a given $x_o$, the integrability condition
(\ref{reflectionless}) determines $K_o$; this condition also determines
that the sign of $x_o$ is given by the sign of $M_f$.
This is an important point. The stable minimum
 $(\Delta_o+)$ always corresponds to
$M_f > 0$, and consequently $x_o >0$. The polaron of the
stable phase {\it decreases} the value of the local dimerization in
the region of the lattice distortion, thus sampling a region of higher
energy density (trans). When $x_o$ becomes large the energy will
{\it grow} linearly with $x_o$.
 This is the ``confining'' mechanism\cite{bk,fbc,ylu} in the stable phase.

Now consider the case in which  $\Delta_o$ is the value at the metastable
minimum $(\Delta_{o-})$. In
this case $M_f <0$ and consequently $x_o <0$. The polaron
in this phase {\it increases} the value of the local dimerization in
the distortion region, thus sampling regions with lower energy density.
Thus for large $\mid x_o \mid$ the energy will {\it decrease} linearly
with  $\mid x_o \mid$ and it becomes favorable for the system to
produce large polarons. But, clearly, there will be an energy barrier to
do so because, for small radius, the elastic energy will gain as will be
 shown below.

iii) The integrability condition (\ref{reflectionless}) yields
non-trivial solutions for $K_o$ only when $2 |x_o| > v_f / |M_f|$.

For $2|x_o| < v_f/ |M_f|$ there is no polaron  solution and the only
available solution is that for constant dimerization.
The physical reason for this is that $v_f/|M_f|$ is the Compton wavelength of
the fermions, and there cannot be bound states localized within a region in
space smaller than the Compton wavelength.

In the other limit
when $|x_o| \rightarrow \infty$, $K_o \rightarrow |M_f|
 / v_f$, the electronic bound states
(see below) merge at the middle of the
gap (zero energy). In this case, the polaron looks like a
widely separated kink-antikink pair, each with a localized
 fermionic zero mode; the two bound states
 correspond to the symmetric and antisymmetric
 combinations of the fermionic ``zero mode'' wave functions, split-off in
energy
because of the (small) overlap of their wave-functions.

It remains to compute the energy of the polaron. There are two
 contributions: the elastic contribution from the term $\Delta^2(x)$ in the
Hamiltonian, and the electronic energy. Due to the reflectionless
nature of the potential, the electronic spectrum is known
exactly\cite{bk,cb}: there are two electron bound states at energies
$ \pm \omega_o =\pm \sqrt{M_f^2-K^2_o v_f^2}$ (the wave
functions of these bound states are localized at $x_{cm}\pm x_o$) and the
conduction and valence continuum states with dispersion $E_c=\pm
\sqrt{k^2 v_f^2+M_f^2}$. The valence and conduction bands are
both depleted by one state per spin degree of freedom, the scattering phase
shifts of the valence
band states are\cite{cb}
\begin{equation}
\delta(k) = 2\tan^{-1}\left[\frac{K_o}{k}\right] \label{phaseshift}
\end{equation}
The form of the bound states and continuum electronic wave functions
has been given exactly by Campbell and Bishop\cite{cb} for the polarons
in the {\it degenerate case}. Their results can be
applied vis a vis to our case because they are properties of the electronic
``Dirac'' equation in the presence of a spatially varying background
configuration whose profile is the same as in the degenerate case.
The reader
is referred to those articles for more details.

The difference in energies between the polaron and the constant dimerization
case is written as
\begin{eqnarray}
\delta E     & = & \delta E_l + \delta E_{el} \label{deltaE}\\
\delta E_l   & = & \frac{1}{\pi v_f \lambda} \int dx
\{\Delta_p^2(x;x_{cm};x_o)-
\Delta_o^2 \} \label{deltaEl} \\
\delta E_{el}& = & (n^+ - n^-)\omega_o -2 \sum_{k} [\omega^p(k)-\omega^o(k)]
\label{deltaEel}
\end{eqnarray}
where we have assumed that the positive and negative energy bound states
are occupied with $n^+$, $n^-$ electrons, respectively ($n^{\pm} = 0,1,2$),
the factor 2 accounts
for the two spin projections and $\omega^{p,o}$ are the energies of the
states in the valence band, with and without the polaron, respectively.
We find
\begin{eqnarray}
\delta E_l & = & \frac{8 \Delta_e K_o x_o-4 K_o v_f}{\pi \lambda}
\label{deltaElfin} \\
\delta E_{el}
           & = & (n^+-n^-)\omega_o+\frac{4}{\pi}K_ov_f+\frac{4}{\pi}K_ov_f
           \left(\frac{\Delta_o}{\lambda M_f}\right) \nonumber \\
           & + & \frac{4}{\pi}\omega_o\left(\frac{\pi}{2}-\tan^{-1}\left[
           \frac{K_ov_f}{\omega_o}\right]\right) \label{deltaEelfin}
\end{eqnarray}

The first term in (\ref{deltaElfin}) has the important physics that we
we were looking for.
Whereas the rest of the
terms in (\ref{deltaElfin},\ref{deltaEelfin}) reach a constant value when
$|x_o| \rightarrow \infty$, it is this first term that dominates the
contribution to the energy in this limit.

As argued above, for the polaron in the
metastable state $x_o<0$, whereas
for the polaron in the stable state $x_o > 0$. Thus, in the metastable
state, as the size of the polaron becomes large, the energy becomes large
and {\it negative}, linearly with the size. In the stable state it becomes
large and positive linearly (this is the confining mechanism found
in\cite{bk,fbc,ylu}).

To compare the energy in the metastable state with the results quoted in
the literature for the stable case, it proves convenient to introduce the
variables $\theta$ and $\gamma$ as
\begin{eqnarray}
K_o v_f  & = & \mid M_f \mid \sin(\theta) \; ; \;
\omega_o  =    \mid M_f \mid \cos(\theta) \; \; ; \; \;
0 \leq \theta \leq \frac{\pi}{2} \label{teta} \\
\gamma   & = & \frac{\Delta_e}{\lambda M_f} \label{gamma}
\end{eqnarray}
 Finally, the energy difference
 in terms of
these variables is given by
\begin{eqnarray}
\delta E & = & \frac{4}{\pi} \mid M_f \mid \left\{
\gamma \tanh^{-1}[\sin(\theta)]+\sin(\theta)-\gamma \sin(\theta)+
\frac{\pi}{4}(n^+-n^-)\cos(\theta)+ \right. \nonumber \\
         &   & \left. \cos(\theta)\left(\frac{\pi}{2}-\theta \right)
\right\} . \label{energyteta}
\end{eqnarray}
It is important to recognize that, for the minimum at the metastable
state, $\gamma <0$.

This expression is similar in form to the one found in references
 (\cite{bk,fbc,ylu,cac}) with
the only difference being the subtleties associated with the signs
for the metastable case.

To contrast the polaron solutions in the metastable state to those
 in the stable state, let us look at the extremum condition obtained
 from the energy function(\ref{energyteta}). The integrability condition
 (\ref{reflectionless}) relates $K_o$ (and $\omega_o$) and phase shifts
 to $x_o$. The value of $x_o$ is obtained from the extremization
 condition of the energy. Because $x_o$ is a monotonically increasing
 function of $\theta$ in the interval $0 \leq \theta \leq \pi/2$, it
 proves more convenient to extremize with respect to $\theta$. We obtain
 the equation
\begin{equation}
\theta-|\gamma| \tan (\theta)=\frac{\pi}{4}(n_+-n_- +2). \label{polsol}
\end{equation}
 For $|\gamma| >1$ the only solution is the trivial one, $\theta =0$,
 however this value of $|\gamma|$ is not within the allowed range of
 parameters that describe a metastable situation.
 For $|\gamma| <1$ there are several interesting possibilities:

i) for $(n_+-n_-)=-2$ there are {\it always} two solutions, one corresponding
to the trivial case $\theta =0$ and another non-trivial solution. This is in
marked contrast with the stable situation for which, at these values  of the
occupation for the bound states,
there is {\it only} the trivial solution. Thus, in the metastable
 case there is always a polaron solution for the ground state
 configuration $n_+ = 0, n_- = -2$. Because the energy difference
 between the constant dimerization and the spatially varying case
 vanishes at $\theta =0$ and grows linearly with negative slope for
 large $|x_o|$ ($\theta \rightarrow \pi/2$), this polaron solution
 corresponds to a {\it maximum} of (\ref{energyteta}) as will be shown
explicitly below.

ii) When $n_+-n_-+2 > 0$ non-trivial solutions are available for a particular
range of parameters and will be studied numerically below. However, we find
that whenever non-trivial solutions are available, they  always appear in
pairs. A similar analysis as presented above reveals that the solution with the
smallest value of $\theta$ corresponds to a minimum and the largest to a
maximum of (\ref{energyteta}). These solutions
correspond to polaron-like configurations in which a lattice
 distortion traps electrons in positive and negative energy bound
 states and are the analog of the polaron solutions found in
 references (\cite{bk,fbc,ylu}) for the stable configuration.

Figures (2.a,b,c) show the energy as a function of $r_o=M_f x_o/v_f$
  for
 $n_+ - n_- = -2;-1;0$ for $\lambda = 0.4077$,
 $\Delta_e=0.02 \rm{\mbox{ eV }}$. For $n_+ - n_- = 0$, there is no
 minimum or maximum, i.e. there are no solutions to the extremum
 equation
 (\ref{polsol}); we will refer to this case as the
 ``dissociation curve'',
 for reasons that will become clear later. Similar curves are
 obtained
 for $\Delta_e= 0.04 \rm{\mbox{ eV }}$, where again there are
 polaronic
 solutions only for $n_+-n_-=-2;-1$ and dissociation curves for
 $n_+-n_- \geq 0$ . For $\Delta_e= 0.06 \rm{\mbox{ eV }}$
 only for the ground state configuration $n_+ - n_- = -2$ there is
 an extremum ``polaron'' solution corresponding to a maximum of the
 energy, for all other values of $n_+ - n_- $
we find dissociation curves without extrema. From the figures, one
finds that in all cases the typical size of the polaron solutions is
$2|x_o| \approx 20-40 \AA$.

\section{Decay of the Metastable State: Thermal Activation}

We are now in condition to study the mechanism that leads to the decay
 of the metastable state. The polaron configurations corresponding to
 the maximum of the energy (\ref{energyteta}) are identified as
 Langer's critical droplets\cite{langer,affleck} or the
 ``transition state''\cite{borkovec}. These configurations correspond
 to a saddle point of the energy functional in the multidimensional
 configuration space. The ``critical
droplet'' is determined by the value $x_o=x_o^*$ at which the energy
 (\ref{energyteta}) is
 a {\it maximum} (the maxima in figures 2.a-c).
 If we consider small fluctuations around the critical droplet
 (transition state) configuration, we find that there is one
 ``zero mode'' corresponding to translations, one unstable mode
 corresponding to dilation of the radius of the droplet and, presumably,
 infinitely many perpendicular directions with positive real frequencies,
 for small oscillations around the droplet configuration. The presence
  of the translational and unstable mode is easy to understand.
Because of translational invariance the energy does not depend on
 the position of the polaron, that is, the coordinate $x_{cm}$ is cyclic.
 Then, a fluctuation around the polaron solution of the form
\begin{equation}
\delta_o(x) = a_o \frac{\partial \Delta_p(x)}{\partial x}
\end{equation}
with $a_o$ constant and small, corresponds to a shift of the polaron position
\begin{equation}
\Delta_p(x;x_{cm};x_o)+\delta_o(x) \approx \Delta_p(x;x_{cm}-a_o;x_o)
\label{zeromode}
\end{equation}
but, by translational invariance, the energy functional is invariant
 under such a shift. Thus, the function
\begin{equation}
f_o(x;x_{cm};x_o) =  \frac{\partial \Delta_p(x)}{\partial x}
\end{equation}
and the collective coordinate $a_o$ determine a ``flat'' direction in
 functional space associated with the ``zero mode''.
 Notice that the function $f_o(x)$ is antisymmetric and has one node.

Now consider a small fluctuation of the polaron solution
 around the value of $x_o=x_o^*$  with $2x_o^*$  the value of
 the radius of the polaron corresponding to the {\it maximum} of
 the energy function (\ref{energyteta}). This fluctuation is
 determined by the function
\begin{equation}
\delta_1(x) = a_1 \frac{\partial \Delta_p(x)}{\partial x_o}|_{x_o=x_o^*}
\end{equation}

For small $a_1$
\begin{equation}
\Delta_p(x;x_{cm};x_o^*)+\delta_1(x) \approx \Delta_p(x;x_{cm};x_o^*+a_1)
\end{equation}
Since $x_o^*$ corresponds to a maximum of the energy functional,
 the coordinate $a_1$ determines (locally) an unstable direction
 in function space around the extremum solution.
It is important to notice that the function $f_1(x)$ is symmetric
(nodeless) and is thus orthogonal to $f_o(x)$.

Generalizing Holstein's\cite{holstein,raja,neto}
 treatment of the large polaron to incorporate $x_o$ as a
 collective coordinate, we write the quantum expansion
 around the polaron solution as
\begin{equation}
\hat{\Delta}(x)= \Delta_p(x-\hat{x}_{cm};\hat{x}_o)+ \sum_{l>1}a_l
f_l(x-\hat{x}_{cm};\hat{x}_o) \label{deltaexpan}
\end{equation}
where the functions $f_l(x)$ are constrained to be orthogonal to $f_o(x)$ and
$f_1(x)$ and correspond to the stable modes of oscillations around the polaron
solutions.

{}From the Hamiltonian (\ref{finham}) we find
\begin{equation}
H= \frac{M_o^2}{2}\dot{x}_{cm}^2+
\frac{M_1^2}{2}\dot{x}_{o}^2+E(x_o)+ \cdots \label{collcorham}
\end{equation}
with
\begin{eqnarray}
M_o^2 & = & \frac{\int dx \left[\partial \Delta_p / \partial x \right]^2}
{\lambda \pi v_f \omega_Q^2} \label{masszeromod} \\
M_1^2 & = & \frac{\int dx \left[\partial \Delta_p /
\partial x_o \right]^2}{\lambda \pi v_f \omega_Q^2} \label{massunstmod}
\end{eqnarray}
and
where the dots represent the coordinates associated with other functional
directions and interactions,
 and $E(x_o)$ is the Born-Oppenheimer energy (\ref{energyteta})
 in terms of $\theta(x_o)$. The masses $M_o$ and $M_1$ are numerically
 found to  have a very weak
dependence on $x_o$ for values of $x_o$ in the region of interest (near
the maxima and minima). We find
\begin{equation}
M_o \approx M_1 \approx 4.2 (\rm{\mbox{ eV }} \AA^2)^{-1} \label{masses}
\end{equation}

Following Langer\cite{langer} and Affleck\cite{affleck} (see
 also\cite{borkovec}) the decay
 rate of the metastable state is obtained as
\begin{equation}
\Gamma = \frac{\Omega}{2\pi}Im F
\label{rateform}
\end{equation}
where Im F is the imaginary part of the analytically
 continued free energy computed in the saddle point
 approximation around the polaron solution corresponding
 to the maximum of the energy function (i.e. the transition state) and $\Omega$
is the (unstable) frequency at the top of the barrier
 along the functional direction determined by dilation of the radius
 of the polaron.
Near the maximum of the energy function, along the unstable direction, the
Hamiltonian becomes
\begin{equation}
H= E(x_o^*)+ \frac{M_o^2}{2}\dot{x}_{cm}^2+
\frac{M_1^2}{2}\dot{x}_{o}^2-\frac{M_1}{2}\Omega^2 (x_o-x_o^*)^2
  +\cdots \label{collcorham2}
\end{equation}
where $E(x_o^*)$ is the energy at the top of the barrier (the maximum)
and the dots again stand for the stable modes and possible interactions.
We find numerically the following values for the energies at the maxima
 ($E(x_o^*)$), minima ($E_{min}$), bound state energies
$\omega_{o\pm}$ at the maxima ($+$)  and minima ($-$) respectively,
 and unstable frequency $\Omega$ at the top of the barrier
 ($\Delta n = n_+ - n_-$).

\begin{equation}
\begin{array}{|r|c|c|c|c|c|l|}\hline
\Delta_e (eV) & \Delta n & E(x_o^*) (eV) & E_{min}(eV) & \omega_{o+}(eV) &
\omega_{o-}(eV) & \Omega (eV)  \\ \hline
0.02          & -2       & 0.792         & ----        & 0.032           &
----           & 0.031        \\ \hline
0.02          & -1       & 0.838         & 0.722       & 0.070           &
0.535           & 0.103        \\ \hline \label{tabletwo}
0.04          & -2       & 0.573         & ----        & 0.066           & ----
           & 0.077        \\ \hline
0.04          & -1       & 0.672         & 0.655       & 0.171           &
0.425           & 0.070        \\ \hline
0.06          & -2       & 0.393         & ----        & 0.102           & ----
           & 0.149        \\  \hline
\end{array}
\end{equation}
Thus, we see that the unstable frequencies at the top of the barriers are
typically of the same order of magnitude as the bare phonon frequencies.

The decay rate is thus obtained by calculating the partition function at
the saddle point, the unstable mode is treated via an
 analytic continuation,
 normalized to the partition function in the metastable well. The zero
mode leads to a volume dependence (L)\cite{langer} and we find
\begin{equation}
\frac{\Gamma}{L} = \frac{\Omega}{2\pi\sin(\Omega/k_BT)}\left[M_o k_B
T\right]^{1/2}
\exp\left[-\frac{E(x_o^*)}{k_BT}\right] [K]. \label{temprate}
\end{equation}
[K] is the ratio of the square root of the determinants for the
stable modes around the extremum polaron solution and the constant
dimerization solution for the metastable well at finite temperature.
 The computation of this
ratio of determinants is beyond our capabilities, it is a dimensionless
number, presumably of order one, as the frequency scales in both cases are
of the same order of magnitude.
The singularities in (\ref{temprate}) at $T=\Omega/ 2 \pi k_B n$
 have been
discussed by Wolynes\cite{wolynes} and the reader is referred to that
article for details. In our case for $n \neq 0$ these values of the
temperature are below the crossover value for thermal activation to
quantum nucleation (see discussion below)
 and the rate given by (\ref{temprate}) is no longer
applicable.
As shown in the table above, $0.4 eV \leq E(x_o^*) \leq 0.8 eV$ thus
for temperatures $T \approx 100-200 K$ the Kramers-Arrenhius factor in
(\ref{temprate}) is fairly large and the {\it lowest energy state}
in the metastable phase (trans), the constant dimerization, is fairly
long-lived.

\subsection{\bf Induced decay}

The lowest energy configuration in the metastable state is that for
 constant
dimerization with the valence band completely
filled.
 Upon doping with electrons (or holes) the metastable phase,
 the energetically most favorable configuration corresponds to the
formation of a polaron
in the Born-Oppenheimer surface corresponding to $n_+ - n_+ = -1$ with
either one electron in the bound state at $+\omega_o$ and two electrons
in the bound state at $-\omega_o$ (electron polaron) or one electron
in the bound state at  $-\omega_o$ (hole polaron).
 On this energy surface, the lowest
energy polaron configuration corresponds to the minimum (available for
$\Delta_e \leq 0.06 \rm{\mbox{ eV }}$).

{}From the table above, we see that the difference in energy between the
minimum and maximum polaron configuration on these surfaces ($E(x_o^*)-
E_{min}$) is typically an {\it order of magnitude smaller}. In this
situation, the activation energy barriers are of order $\Delta E \approx
0.02 - 0.1 \rm{\mbox{ eV }}$, and the decay rate is dramatically
 enhanced.

 Upon further doping, the Born-Oppenheimer surface may change to
$n_+ - n_- \geq 0$ and the dissociation curve is reached for almost all
phenomenologically available values of $\Delta_e$.
In particular, this is always the case for bipolarons $n_+=n_-$ for the
range of parameters consistent with the phenomenology of
 cis-polyacetylene; this is another major difference with the polarons in the
stable phase.
 In this case, there
are no barriers and the system decays instantaneously. The decay
 process, in
this situation, cannot be studied under the assumptions of
quasi-equilibrium implied in the treatment of metastable decay and a
time dependent non-equilibrium treatment would have to be used.

The first excited and dissociation curves may also be reached by absorption of
photons
of energy $\hbar \nu = |M_f|+\omega_o$ and electronic transitions to
the bound state with energy $+\omega_o$.
An absorption peak in this energy
 range will
be the telltale of electron bound states, just as absorption peaks in
the degenerate case reveal the existence of electrons bound to
solitons\cite{blanchet}. Since $\omega_o$ is typically very small,
these states may appear as mid-gap states at $\hbar \nu \approx 1 eV$.
 However, there is no room for
confusion with bound states on solitons, because solitons are not
available in the non-degenerate case.

\section{\bf Decay via Quantum Tunneling}

At very low temperatures (to be quantified later), we expect that the
metastable state will decay via quantum tunneling. In order to understand
quantum tunneling in multidimensional space, one must search for the
configurations that extremize the {\it classical action} in imaginary
(euclidean) time. These configurations constitute the ``most probable
escape path''\cite{bitar,sethna}, and are solutions to the classical
equations of motion in euclidean time\cite{volo}, dubbed
``bounces''\cite{coleman1,callancoleman,coleman2,stone,schmid}.

In terms of the collective coordinate $x_o$ representing the radius of
the polaron, this ``most probable escape path'' or ``bounce''
 corresponds to the
classical trajectory in the {\it inverted} potential $-E(x_o)$ between
the two classical turning points corresponding to the energy of the
metastable state. For the ground state Born-Oppenheimer curves
 ($n_+ - n_-=-2$) this energy is zero, whereas for the higher energy
surfaces ($n_+ - n_-=-1$) the energy is that of the minimum and marked
with a dashed line in figure (2.b). This approach was previously used
 by Kivelson and Sethna\cite{kivset} to study photoinduced soliton-pair
production.

To exponential accuracy in the semiclassical WKB approximation, the
decay rate  per unit length is given by
\begin{equation}
\frac{\Gamma}{L} \approx \exp\left[-2\frac{S_o}{\hbar}\right] \label{wkbdecay}
\end{equation}
with $S_o$ the action of the classical trajectory in euclidean time
 at energy E between
the two classical turning points $x_-$; $x_+$:
\begin{equation}
S_o= \int^{x_+}_{x_-}dx_o \sqrt{2M_o[E(x_o)-E]} \label{action}
\end{equation}
Again, the total decay rate will have  prefactors in front of the WKB
exponential in (\ref{wkbdecay}); this prefactor involves the ratio of
the determinants of the quadratic fluctuations.
 We are unable at the moment to calculate the prefactor and content
ourselves with an estimate of the exponential, which is the leading term
in the semiclassical approximation.

We can provide a rough estimate of the crossover temperature between
thermal activation and quantum nucleation by comparing the WKB factor
to the Arrhenius-Kramers activation factor, thus obtaining the approximate
estimate for the crossover temperature
\begin{equation}
k_B T_c \approx \frac{\hbar \Delta E}{2S_o} \label{crosstemp}
\end{equation}
where $\Delta E$ is the activation barrier ($E(x_o)$ for the ground
state or $E(x_o)-E_{min}$ for the higher energy surface).
The action $S_o$ is calculated numerically and we find:

\begin{equation}
\begin{array}{|r|c|c|l|}\hline
\Delta_e (eV) & \Delta n & \frac{S_o}{\hbar} & k_B T_c (eV)   \\ \hline
0.02          & -2       & 102.7             & 0.004          \\ \hline
\label{tablethree}
0.02          & -1       & 11.3              & 0.005          \\ \hline
0.04          & -2       & 41.3              & 0.007          \\ \hline
0.04          & -1       & 1.4               & 0.006          \\ \hline
0.06          & -2       & 21.0              & 0.009          \\ \hline
\end{array}
\end{equation}

Thus, we see that, typically, the crossover temperature is of the order of
$T_c \approx 40-100 K$ depending on the value of the intrinsic
 dimerization, and may be amenable to experimental realization.

Another (but related) criterion for the crossover temperature is obtained
from the quasiequilibrium approach\cite{schmid}. It is based on the
properties of the ``bounce'' solution in Euclidean time corresponding
to a classical trajectory with period $\beta \hbar$
in the inverted potential well. These trajectories give the leading
semiclassical contribution to the equilibrium partition function.
 The smallest period for trajectories in the
inverted potential corresponds to the harmonic oscillations with frequency
$\Omega$ at the bottom of the (inverted)
 well (see equation \ref{collcorham2}).
Thus the maximum temperature for which there are ``bounce'' solutions in
Euclidean time is given by
\begin{equation}
\beta \hbar = \frac{2\pi}{\Omega}
\end{equation}
leading to an estimate for the crossover temperature\cite{borkovec}
\begin{equation}
T_c = \frac{\hbar \Omega}{2\pi k_B}.
\end{equation}
 With the values of $\Omega$ from (\ref{tabletwo}), we find the estimate
for crossover temperature based on this criterion to be
 $50 K \leq T_c \leq 100 K$ which is consistent with the previous estimate
based on the comparison of the Arrenhius-Kramers and WKB factors.
For $T<T_c$ the decay process will be dominated by quantum nucleation and
will be fairly insensitive to temperature, whereas for $T>T_c$ the decay
will be dominated by thermal activation and the rate will depend strongly on
temperature
via the Arrenhius-Kramers factor in the decay rate.
Again, quantum decay will be accompanied by the presence of electron
bound states at energies $\pm \omega_o$ and may be detected via absorption
peaks, just as in the degenerate case, where the absorption peak at the
middle of the gap signals the presence of solitons.

Quantum nucleation can also be induced by doping, thus moving from one
Born-Oppenheimer surface, for example the ground state, to the next
for which $n_+ - n_- = -1$. We see that the height and width of the
barrier for this surface are much smaller, resulting in a ten-fold reduction
of the WKB action. Thus, the transmission probability, i.e. the nucleation
rate, is greatly enhanced. Upon further doping, the dissociation curve is
reached and there is no quasi-stationary state; one must
resort to a full real time non-equilibrium calculation of the decay
probability in this case.

The decay through quantum nucleation occurs via the spontaneous
 production of a fairly large droplet; this is  a tunneling process.
 The size of the droplet will be
 roughly determined by the value of the coordinate $|x_o|$ at the
 classically forbidden turning point below the barrier, i.e. the ``escape
point''. For the lowest energy
 surface, this value is typically $\approx 40 \AA$ whereas for the first
 excited surface it is about $\approx 20 \AA$. Because of these large
values of the radius of the quantum droplet, the bound states will
 be very close to the middle of the gap and this quantum droplet looks
 like a kink-antikink pair widely separated.

The telltale signal for the decay of the metastable (trans-cisoid) state,
either via thermal activation or quantum nucleation,
will be electron bound states with energies very near the middle of the
gap (because for the typical sizes of the polarons, the energy $\omega_o$
is very small) and detected through absorption peaks at about
$\hbar \nu \approx 1 eV$.
 These peaks could not be confused
with solitons as these topological excitations are not available in the
non-degenerate isomers. Furthermore, the different regimes for thermal
activation and quantum nucleation may be separated by plotting
the logarithm of the nucleation rate versus $1/T$: activated processes
lead to an approximate straight line with negative slope,
 whereas for quantum
nucleation there should be a flat plateau. Our analysis for the crossover
temperature leads to the prediction that such a plot should have
a plateau for temperatures smaller than about $40 K$.

We must say that at the moment we are not aware of experiments that
 have
either looked at or reported on these possibilities.

\section{Conclusions and speculations}

The motivation of this work was to study the mechanism responsible for
the decay of the metastable (trans-cisoid) configuration to the stable
(cis-transoid) isomer in non-degenerate polymers. We have identified the
electron-phonon configurations that play the role of the nucleation
droplets as polarons in the {\it metastable phase}, and have
studied their properties within the range of parameters for the intrinsic
dimerization compatible with a gap of about 2 eV
in the electronic spectrum for the cis-isomer. In the undoped, lowest
energy metastable configuration, typical barrier energies are of the order
of about $0.4-0.8$ eV depending on the intrinsic dimerization parameters.
Upon doping with electrons
or holes, higher energy Born-Oppenheimer surfaces become available and
activation barriers become smaller by almost an order of magnitude; thus,
upon doping, the metastable decay is induced and the rate is enhanced
dramatically. For large doping, dissociation curves become available
without activation barriers; in
this  case a full time dependent non-equilibrium study will be needed
to understand the decay.

We obtained an estimate for the decay rate via thermal
activation by treating the relevant coordinates of the polaron (center
of mass and radius) as collective coordinates and evaluated numerically
the necessary activation barriers and unstable frequencies to obtain
the rate approximately. We also performed a semiclassical WKB calculation to
estimate
 the decay rate via quantum nucleation and to establish the crossover
temperature that separates the regimes between thermal activation and
quantum nucleation. We found these temperatures to be approximately
$T_c \leq 40 K$.

We speculate that metastable isomers in non-degenerate polymers may be
candidates to study mesoscopic quantum tunneling. Typical values for
the sizes of critical droplets (polarons) are $20-40 \AA$. The telltale
signal that the metastable decay is mediated by polarons will be
electron bound states with energies very near the middle of the gap.
These states may be detected via absorption peaks at about 1 eV,
 just as the electron
states bound to solitons in the degenerate case. There is no possibility
to confuse the metastable polarons with solitons as the latter are not
available excitations in the non-degenerate case. Furthermore, our
rough estimate for the crossover temperature should be amenable to
experimental confirmation by plotting the logarithm of the nucleation
 rate versus $1/T$.
In the thermally activated regime this should be almost a straight line
with negative slope, the curve should flatten and reach a plateau at
temperatures $T \leq 40 K$, regime in which the nucleation rate is insensitive
to temperature signaling quantum nucleation.

{\bf Acknowledgements}
D. B. would like to thank M. Stone, D. Jasnow, R. Coalson, P. Goldbart and W.
Goldburg for stimulating conversations and suggestions.
D. B. acknowledges support from the National Science Foundation through
grant No: PHY-9302534, and for support through a binational collaboration
from the International Programs at N.S.F. through grant No: INT-9016254.
D. B. would also like to thank the hospitality at the Institute for
Theoretical Physics at Santa Barbara where part of this work was done and
partially supported by N.S.F. grant No: PHY-89-04035. He also acknowledges a
grant from the Pittsburgh Supercomputer Center No: PHY940005P.
 D. B. and C.A.A.C. would like to thank F. Moraes for fruitful conversations.
C.A.A.C. would like to thank CNPq for support through a binational
collaboration
and E.S.F. would like to thank CNPq for a Master's scholarship.

\newpage

\underline{\bf Figure Captions:}

\vspace{3mm}

{\bf Figure 1}: $\frac{E(\Delta_o)}{N}$ (eV) vs. $\Delta_o$ (eV)
 for $\lambda = 0.4077$;  $\Delta_e = 0.02 \rm{\mbox{ eV }}$.

\vspace{3mm}

{\bf Figure 2.a}: $\delta E(r_o)$(eV) (equation (\ref{energyteta}))
vs. $r_o = M_f x_o/v_f$, for $n_+ - n_- = -2$.
 The ``bounce'' trajectory
has zero energy.

\vspace{3mm}

{\bf Figure 2.b}: $\delta E(r_o)$(eV) (equation (\ref{energyteta}))
vs. $r_o = M_f x_o/v_f$, for $n_+ - n_- = -1$. The dashed line represents
the energy for the ``bounce'' trajectory.

\vspace{3mm}

{\bf Figure 2.c}: $\delta E(r_o)$(eV) (equation (\ref{energyteta}))
vs. $r_o = M_f x_o/v_f$, for $n_+ - n_- = -2; -1 ; 0$.


\begin{thebibliography}{99}
\bibitem{note} The continuum theory for the degenerate (trans) case
predicts a gap with a small discrepancy with the experimental results, when the
standard parameters\cite{ssh} are used.
\end{thebibliography}
\end{document}